\documentstyle[12pt]{article}

\begin{document}
\setcounter{page}{1} \pagestyle{plain} \vspace{5cm}
\begin{center}
\Large{\bf Crossing the Phantom Divide Line in a DGP-Inspired
$F(R,\phi)$-Gravity}\\
\small
\vspace{1cm} {\bf Kourosh Nozari}\quad and \quad{\bf Mahin Pourghasemi}\\
\vspace{0.5cm} {\it Department of Physics,
Faculty of Basic Sciences,\\
University of Mazandaran,\\
P. O. Box 47416-95447, Babolsar, IRAN\\
knozari@umz.ac.ir}
\end{center}
\vspace{1.5cm}
\begin{abstract}
We study possible crossing of the phantom divide line in a
DGP-inspired $F(R,\phi)$ braneworld scenario where scalar field and
curvature quintessence are treated in a unified framework. With some
specific form of $F(R,\phi)$ and by adopting a suitable ansatz, we
show that there are appropriate regions of the parameters space
which account for late-time acceleration and admit crossing of the
phantom divide line.

{\bf Key Words}: Braneworld Cosmology, DGP Scenario, Dark Energy Models, Late-time Acceleration\\
{\bf PACS}: 04.50.+h,\, 98.80.-k
\end{abstract}
\newpage
\section{Introduction}
Based on several astronomical evidences, our universe is currently
in a period of positively accelerated expansion [1]. It is possible
to interpret this late-time acceleration based on yet unknown
component called dark energy in literature ( see for instance [2]
with a comprehensive list of references therein). Also, it has been
shown that such an accelerated expansion could be the result of a
modification to the Einstein-Hilbert action ( for a recent review
see [3]). On the other hand, DGP braneworld scenario has the
capability to interpret this late-time acceleration via leakage of
gravity to extra dimension in its self-accelerating branch [4]. For
the first alternative, the simplest candidate for dark energy is the
cosmological constant itself. However, it suffers from serious
problems such as a huge amount of fine-tuning [2,5,6]. Within dark
energy viewpoint, for a dark energy component, say $X$, its equation
of state or equivalently equation of state parameter
$\omega_{X}=\frac{p_{X}}{\rho_{X}}$, determines both gravitational
properties and evolution of the dark energy. Recent constraints on
equation of state of dark energy indicate that $p_{X}\approx -
\rho_{X}$ and even that $\omega_{X}< -1$. When we consider
background evolution, there is no problem with $\omega_{X}< -1$.
However, there will be apparent divergencies in perturbation when
one crosses the phantom divide line, $\omega_{X}= -1$. Accordingly,
crossing of the phantom divide line, $\omega_{X}= -1$, provides a
suitable basis to test alternative theories of gravity or higher
dimensional models which can give rise to an effective phantom
energy [6]. With these motivations, the issue of phantom divide line
crossing has been investigated extensively in recent years ( see [6]
and references therein). In principle, crossing of phantom divide
line by the dark energy equation of state parameter at recent
red-shifts, has two possible cosmological implications: either the
dark energy consists of multiple components with at least one
non-canonical phantom component or general relativity needs to be
extended to a more general theory on cosmological scales. Both of
these conjectures have been studied extensively. For instance,
curvature quintessence [7] is a fascinating proposal in this
respect. Recently, phantom-like behavior in a brane-world model with
curvature effects and also in dilatonic brane-world scenario with
induced gravity have been reported [8]. Dark energy models with
non-minimally coupled scalar field and other extensions of
scalar-tensor theories have been studied widely some of which can be
found in reference [9]. In the spirit of scalar-tensor dark energy
models, our motivation here is to show that a general $F(R,\phi)$ \,
DGP-inspired scenario can account for late-time acceleration and
crossing of phantom divide line in some suitable domains of model
parameters space. To show this feature, first we study cosmological
dynamics of $F(R,\phi)$ \,DGP-inspired scenario briefly. In the
minimal case, motivated by modified theories of gravity, some
authors have included a term of the type $\ell_{0}R^{n}$ in the
action [7,10]. This extension for some values of $n$ has the
capability to explain late-time acceleration of the universe in a
simple manner. It is then natural to extend this scenario to more
general embedding of DGP inspired scenarios. The purpose of this
paper is to perform such a generalization to study both late-time
acceleration and possible crossing of phantom divide line in this
setup. We focus mainly on the potential of the type
$V(\phi)=\lambda\phi^{\eta}$ for non-minimally coupled scalar field.
Our setup for minimal case predicts a power-law acceleration
supporting observed late-time acceleration. In the non-minimal case,
by a suitable choice of non-minimal coupling and scalar field
potential, one obtains accelerated expansion in some specific
regions of parameters space. While a single minimally coupled scalar
field in four dimensions cannot reproduce a crossing of the phantom
divide line for any scalar field potential [11], a non-minimally
coupled scalar field account for such a crossing [6]. In DGP model,
equation of state parameter of dark energy never crosses
$\omega(z)=-1$ line, and universe eventually turns out to be de
Sitter phase. Nevertheless, in this setup if we include a single
scalar field (ordinary or phantom) on the brane, we can show that
equation of state parameter of dark energy can cross phantom divide
line [12]. Crossing of phantom divide line with non-minimally
coupled scalar field on the warped DGP brane has been studied
recently [13]. Our purpose here is to obtain an extension of these
non-minimal dark energy models within DGP-inspired $F(R,\phi)$
braneworld scenario. We use a prime for differentiation with respect
to $R$. An overdot marks differentiation with respect to the brane
time coordinate.

\section{$F(R,\phi)$ DGP-Inspired Gravity }
We start with the following action
\begin{eqnarray}
S=\frac{m_{4}^{3}}{2}\int d^{5}(x)\sqrt{-g} \Re + \int
d^{4}(x)\sqrt{-q}\Big(\frac{m_{3}^{2}}{2}
F(R,\phi)-\frac{1}{2}q^{\mu\nu}\nabla_{\mu}\phi\nabla_{\nu}
\phi-V(\phi)+m_{4}^{3}\overline{K}+L_{m}\Big),
\end{eqnarray}
where the first term shows the usual Einstein-Hilbert action in 5D
bulk with 5D metric denoted by $g_{AB}$ and Ricci scalar denoted by
$\Re$. The second term on the right is a generalization of the
Einstein-Hilbert action induced on the brane. This is an extension
of the scalar-tensor theories in one side and a generalization of
$f(R)$-gravity on the other side. We call this model as $F(R,\phi)$
\,DGP-inspired scenario. $y$ is the coordinate of the fifth
dimension and we suppose that brane is located at $y=0$ .
$q_{\mu\nu}$ is induced metric on the brane which is connected to
$g_{AB}$ via $q_{\mu\nu} =
\delta_{\mu}\,^{A}\delta_{\nu}\,^{B}g_{AB}$. $\overline{K}$~ is the
trace of the mean extrinsic curvature of the brane  defined as
follows
\begin{eqnarray}\overline{K}_{\mu\nu}=\frac{1}{2}\lim_{\epsilon\longrightarrow+0}\bigg([K_{\mu\nu}]_{y=-\epsilon}
+[K_{\mu\nu}]_{y=+\epsilon}\bigg)
\end{eqnarray}
We denote matter field Lagrangian by $L_{m}(q_{\mu\nu},\psi)$ with
the following energy-momentum tensor
\begin{eqnarray}T_{\mu\nu}=-2\frac{\delta L_{m}}{\delta
q^{\mu\nu}}+q^{\mu\nu}L_{m}.
\end{eqnarray}
The pure scalar field lagrangian is $L_{\phi}=\frac{1}{2}q^{\mu\nu}
\nabla _{\mu}\phi\nabla_{\nu}\phi-V(\phi)$ which gives the following
energy-momentum tensor
\begin{eqnarray}\top_{\mu\nu}=\nabla_\mu  {\phi}  \nabla_{\nu} \phi -
\frac{1}{2} q_{\mu\nu}(\nabla \phi)^{2}-q_{\mu\nu}V(\phi)
\end{eqnarray}
The field equations resulting from this action are given as follows
\begin{equation}
\frac{m_{4}^{3}}{F'(R,\phi)}\bigg(\Re_{AB}-\frac{1}{2} g_{AB}
\Re \bigg)+m_{3}^{2}
 \delta_{A}\,^{\mu}\delta_{B}\,^{\nu} \Big(R_{\mu\nu}-
\frac{1}{2}q_{\mu\nu}R\Big)\delta(y)=
\delta_{A}\,^{\mu}\delta_{B}\,^{\nu}(\hat{T}_{\mu\nu}+\hat{\top}_{\mu\nu}+
T_{\mu\nu}^{curv})\delta(y).
\end{equation}
In this relation $\hat{T}_{\mu\nu}=\frac{T_{\mu\nu}}{F'(R,\phi)}$
where $T_{\mu\nu}$ is the energy-momentum tensor in matter frame and
$\hat{\top}_{\mu\nu}=\frac{\top_{\mu\nu}}{F'(R,\phi)}$. Also,
$T_{\mu\nu}^{curv}$ is defined as follows
\begin{equation}
T_{\mu\nu}^{curv}=\frac{m_{3}^{2}}{F'(R,\phi)}\bigg[
\frac{1}{2}q_{\mu\nu}\bigg(F(R,\phi)-RF'(R,\phi)\bigg
)+\bigg(F'(R,\phi)\bigg)^{;\alpha\beta}
\bigg(q_{\mu\alpha}q_{\nu\beta}-q_{\mu\nu}q_{\alpha\beta}\bigg)\bigg].
\end{equation}
In the bulk, $T_{AB}=0$  and therefore
\begin{eqnarray}G_{AB}=\Re_{AB}-\frac{1}{2}g_{AB}\Re=0
\end{eqnarray}
and on the brane we have
\begin{eqnarray}G_{\mu\nu}=R_{\mu\nu}-\frac{1}{2}q_{\mu\nu}R=
\frac{\tau_{\mu\nu}}{m_{3}^{2}},
\end{eqnarray}
where $ \tau_{\mu\nu}= \hat{T}_{\mu\nu}+\hat{\top}_{\mu\nu}+
T_{\mu\nu}^{curv}$. The corresponding junction conditions relating
quantities on the brane are as follows
$$\lim_{\epsilon\longrightarrow+0}[K_{\mu\nu}]_{y=-\epsilon}^{y=+\epsilon}=
\frac{F'(R,\phi)}{m_{4}^{3}}\bigg[\tau_{\mu\nu}-\frac{1}{3}q_{\mu\nu}q^{\alpha\beta}
\tau_{\alpha\beta}\bigg]_{y=0}-$$\begin{eqnarray}\frac{m_{3}^{2}}{m_{4}^{3}}F'(R,\phi)
\bigg[R_{\mu\nu}-\frac{1}{6}q_{\mu\nu}q^{\alpha\beta}R_{\alpha\beta}\bigg]_{y=0}\end{eqnarray}
A detailed study of weak field limit of this scenario within
harmonic gauge on the longitudinal coordinates and using Green's
method to find gravitational potential, leads us to a modified
(effective) cross-over distance in this set-up as follows ( see [14]
for details of a similar argument)
\begin{eqnarray}\ell_{F}=\frac{m_{3}^{2}}{2m_{4}^{3}}
\bigg(\frac{dF}{dR}\bigg)=\bigg(\frac{dF}{dR}\bigg)\ell_{DGP},
\end{eqnarray}
where $\ell_{DGP}=\frac{m_{3}^{2}}{2m_{4}^{3}}.$ The gravitational
potential in this scenario takes the following forms in two
different extreme:\\
For~ $r\ll \ell_{F}$
\begin{eqnarray}U(r)=-\frac{(M_{\psi}+M_{\phi})}{6\pi m_{3}^{2} F'(R,\phi)r}
\bigg[1+(\gamma-\frac{2}{\pi})\frac{r}{\ell_{F}}+\frac{r}{\ell_{F}}
\ln(\frac{r}{\ell_{F}})+O
 (\frac{r^{2}}{\ell_{F}^{2}})\bigg]
\end{eqnarray} and for ~$r\gg\ell_{F}$
\begin{eqnarray}
U(r)=-\frac{(M_{\psi}+M_{\phi})}{6\pi^{2}m_{4}^{3}r^{2}}
\bigg[1-\frac{2\ell_{F}^{2}}{r^{2}}+O
(\frac{\ell_{F}^{4}}{r^{4}})\bigg].
\end{eqnarray}
$M_{\phi}$ is effective mass of the scalar field $\phi$ which is
non-minimally coupled to induced gravity, $M_{\psi}$ collectively
shows the effective mass of other material fields minimally coupled
to gravity and $\gamma=0.577$ is Euler's constant. The most
important feature of this $F(R,\phi)$ \,DGP-inspired scenario is the
fact that now cross-over scale is explicitly related to the induced
curvature on the brane and non-minimal coupling of scalar field with
this curvature. Since the dynamics of scalar field $\phi$ is
described by $\nabla^{\mu}\phi\nabla_{\mu}\phi-\frac{dV}{d\phi}+
\frac{m_{3}^{2}}{2}\frac{dF}{d\phi}=0$, one may explain this result
as a spacetime variation of the Newton's constant. In this
viewpoint, when $F(R,\phi)$ varies from point to point on DGP brane,
the crossover scale takes different values. Recent best-fit
crossover scale is given by \,$\ell_{DGP}=1.26H_{0}^{-1}$\, [15]
where \,$H_{0}^{-1}\sim 3000 Mpc $. Since
$\frac{dF}{dR}=\frac{2m_{4}^{3}\ell_{F}}{m_{3}^{2}}$, if we choose
\,$\ell_{F}\sim 1.26H_{0}^{-1}$\, it is possible to constraint this
$F(R,\phi)$ scenario to be consistent with observational data. For
instance, if we set $F(R,\phi)= \alpha(\phi)R$ it is easy to show
that $\alpha(\phi)$ can attain the value of $\alpha(\phi)\sim
2.52\times 10^{-30}H_{0}^{-1}$ where we have assumed
\,$m_{4}\sim10^{3} GeV $ and $m_{3}\sim10^{18} GeV $. This argument
has the potential to be used as a mechanism for obtaining reliable
form of $F(R,\phi)$ in cosmological context.

\section{Cosmological Implications of the Model}
Embedding of FRW cosmology in DGP scenario (which deviates from
general relativity at large distances) is possible in the sense that
this model accounts for cosmological equations of motion at any
distance scale on brane with any function of Ricci scalar. It is
well-known that original DGP model explains late-time acceleration
via leakage of gravity to extra dimension. On the other hand, in
this model equation of state parameter of dark energy never crosses
the phantom divide line. Extension of DGP scenario to more
generalization of the brane action may result in several new
implications on cosmological ground. In this respect, it is
interesting to know whether DGP-inspired $F(R, \phi)$ gravity can
account for late-time acceleration and especially phantom divide
line crossing. With this motivation, in this section we investigate
cosmological implications of our setup focusing firstly on the
late-time acceleration. We assume the following line element
\begin{eqnarray}ds^{2}=q_{\mu\nu}dx^{\mu}dx^{\nu}+b^{2}(y,t)dy^{2}=-n^{2}(y,t)
dt^{2}+a^{2}(y,t)
 \gamma_{ij}dx^{i}dx^{j}+b^{2}(y,t)dy^{2}
\end{eqnarray}
where $\gamma_{ij}$ is a maximally symmetric 3-dimensional metric
defined as $\gamma_{ij}=\delta_{ij}+k\frac{x_{i}x_{j}}{1-kr^{2}}$
where \,$ k=-1,0,1 $ \, parameterizes the spatial curvature and
$r^{2}=x_{i}x^{i}$. We assume that the scalar field $\phi$ depends
only on the cosmic time of the brane. Choosing a Gaussian normal
coordinate system so that  $b^{2}(y,t)=1$, non-vanishing components
of Einstein's tensor in the bulk plus junction conditions on the
brane defined as
$$\lim_{\epsilon\longrightarrow+0}[\partial_{y}n]_{y=-\epsilon}^{y=+\epsilon}(t)=
\frac{2nm_{3}^{2}}{m_{4}^{3}}\bigg[
\Big(\frac{dF}{dR}\Big)\Big(\frac{\ddot{a}}{n^{2}a}-\frac{\dot{a}^{2}}{2n^{2}a^{2}}
-\frac{\dot{n}\dot{a}}{n^{3}a}-\frac{k}{2a^{2}}\Big)\bigg]_{y=0}$$
\begin{eqnarray}
+\frac
{n}{3m_{4}^{3}}\bigg[\Big(\frac{dF}{dR}\Big)\Big(2\rho^{(tot)}+3p^{(tot)}\Big)
\bigg]_{y=0},\end{eqnarray}
\begin{eqnarray}\lim_{\epsilon\longrightarrow+0}[\partial_{y}a]_{y=-\epsilon}^{y=+\epsilon}(t)=
\frac{m_{3}^{2}}{m_{4}^{3}}\bigg[\Big(\frac{dF}{dR}\Big)
\Big(\frac{\dot{a}^{2}}{n^{2}a}+\frac{k}{a}\Big)\bigg]_{y=0}-
\bigg[\bigg(\frac{dF}{dR}\bigg)\frac{\rho^{(tot)}a}{3m_{4}^{3}}\bigg]_{y=0}
\end{eqnarray}
yield the following generalization of Friedmann equation for
cosmological dynamics on the brane ( see [14] for machinery of
calculations for a simple case)
\begin{eqnarray}
H^2+\frac{k}{a^2}=\frac{1}{3m_{3}^2F'(R,\phi)}\Bigg(\rho^{tot}
+\rho_{0}\bigg[1+\epsilon\sqrt{1+\frac{2}{\rho_{0}}\bigg[\rho^{tot}-
\frac{m_{3}^{2}F'(R,\phi)\varepsilon_{0}}{a^4}\bigg]}\bigg]\Bigg)
\end{eqnarray}
where $\epsilon=\pm1$ shows two different embedding of the brane, \,
$\rho_{0}=\frac{6m_{4}^{6}}{m_{3}^{2}F'(R,\phi)}$\, and
\,$\varepsilon_{0}=3\Big(\frac{\dot{a}^{2}}{n^{2}}-a'^{2}+k\Big)a^{2}$\,
is a constant ( with $a'\equiv\frac{da}{dy}$). Total energy density
and pressure are defined as
~$\rho^{(tot)}=\hat{\rho}+\rho_{\phi}+\rho^{curv}$,
~$p^{(tot)}=\hat{p}+p_{\phi}+p^{curv}$~  respectively. The ordinary
matter on the brane has a perfect fluid form with energy density
$\hat{\rho}$ and pressure $\hat{p}$, while the energy density and
pressure corresponding to non-minimally coupled scalar field and
also those related to curvature are given as follows
\begin{eqnarray}\rho_{\varphi}=\bigg[\frac{1}{2}\dot{\phi}^{2}+n^{2}V(\phi)-
6\frac{dF}{d\phi}H\dot{\phi}\bigg]_{y=0},\end{eqnarray}
\begin{eqnarray}p_{\phi}=\bigg[\frac{1}{2n^{2}}\dot{\phi}^{2}-V(\phi)+\frac{2}{n^{2}}
\frac{dF}{d\phi}(\ddot{\phi}-\frac{\dot{n}}{n}\dot{\phi})+
4\frac{dF}{d\phi}\frac{H}{n^{2}}\dot{\phi}+\frac{2}{n^{2}}
\frac{d^{2}F}{d\phi^{2}}\dot{\phi}^{2}\bigg]_{y=0}.\end{eqnarray}
also
\begin{eqnarray}\rho^{curv}=\frac{m_{3}^{2}}{F'(R,\phi)}\bigg(\frac{1}{2}
\bigg[F(R,\phi)-R F'(R,\phi)\bigg] -3\dot{R}H F''(R,\phi) \bigg),
\end{eqnarray}
\begin{eqnarray}p^{curv}=\frac{m_{3}^{2}}{F'(R,\phi)}\bigg({2\dot{R}H
F''(R,\phi)+\ddot{R}F''(R,\phi)
+\dot{R}^{2}F'''(R,\phi)-\frac{1}{2}\Big[ F(R,\phi)-R
F'(R,\phi)\Big]}\bigg).\end{eqnarray} where
$H=\frac{\dot{a}(0,t)}{a(0,t)}$~ is the Hubble parameter. Ricci
scalar on the brane is given by
$$R=3\frac{k}{a^2}+\frac{1}{n^{2}}\bigg[6\frac{\ddot{a}}{a}+
6\Big(\frac{\dot{a}}{a}\Big)^{2}-6\frac{\dot{a}}{a}\frac{\dot{n}}{n}\bigg].$$
In this setup, non-minimal coupling of scalar field and induced
gravity leads to no-conservation of effective energy density on the
brane
\begin{eqnarray}\dot{\rho}^{(tot)}+3H\Big(\rho^{(tot)}+
p^{(tot)}\Big)=6\Big(\frac{dF}{d\phi}\Big)\dot{\phi}(H^{2}+
\frac{k}{a^{2}}).\end{eqnarray} It is easy to show that for a
minimally coupled scalar field on the brane, this setup yields a
late-time accelerating universe in a fascinating manner [10]. The
evolution of the scalar field on the brane for a spatially flat FRW
geometry, $(k=0)$, is described by the following equations
\begin{eqnarray}H^{2}=\frac{1}{3m_{3}
^{2}F'(R,\phi)}\rho^{(tot)}\end{eqnarray} and
\begin{eqnarray}\ddot{\phi}+3H\dot{\phi}+\frac{d V(\phi)}{d\phi}
=\frac{dF(R,\phi)}{d\phi}.\end{eqnarray} and finally
\begin{eqnarray}
\frac{\ddot{a}(0,t)}{a(0,t)}=-\frac{1}{6m_{3}^{2}F'(R,\phi)}\bigg[\rho^{tot}+3p^{tot}\bigg].
\end{eqnarray}
These equations show that essentially embedding of FRW cosmology in
DGP setup is possible. Now, after a brief study of gravitational and
cosmological implications of our setup, we investigate late-time
acceleration and crossing of the phantom divide line in this setup.

\subsection{Accelerated Expansion}
Standard DGP model itself has the capability to explain the
late-time acceleration of the universe via leakage of gravity to
extra dimensions without any additional mechanism [4].  In our
$F(R,\phi)$ DGP inspired model, it is interesting to see whether
there is any room for explanation of this late time acceleration.
The viability of this question lies in the fact that generally with
non-minimal coupling it is harder to achieve accelerated expansion
[16]. We first obtain a necessary condition for the acceleration of
the universe in $F(R,\phi)$ DGP-inspired model. Then we use a simple
ansatz to clarify our general equations. Suppose that scalar field
is the only source of matter on the brane so that $\hat{\rho}=0$ for
other matter candidates. The necessary condition for acceleration of
the universe is \,$( \rho_{\phi}+\rho^{curv})+3(p_{\phi}+p^{curv})<0
$ because in our $F(R,\phi)$ setup scalar field and curvature are
correlated. Therefore we obtain
$$2\bigg(1+3\frac{d^{2}F}{d\phi^{2}}\bigg)\dot{\phi}^{2}+
6\Big(\frac{dF}{d\phi}\Big)\Big(H\dot{\phi}+\ddot{\phi}\Big)+\frac{3m_{3}^{2}}
{F'(R,\phi)}\bigg[\dot{R} ^{2}F'''(R,\phi)+
F''(R,\phi)\Big(\dot{R}H+\ddot{R}\Big)\bigg]-$$$$
m_{3}^{2}\bigg(\frac{F(R,\phi)}{F'(R,\phi)} -R\bigg)-2V(\phi)<0.
$$
Using the Klein-Gordon equation (23), this relation can be rewritten
as follows
$$2\bigg(1+3\frac{d^{2}F}{d\phi^{2}}\bigg)\dot{\phi}^{2}+6\Big(\frac{dF}{d\phi}\Big)^{2}-
6\Big(\frac{dF}{d\phi}\Big)\Big(2H\dot{\phi}+\frac{dV}{d\phi}\Big)+\frac{3m_{3}^{2}}
{F'(R,\phi)}\bigg[\dot{R} ^{2}F'''(R,\phi)+
F''(R,\phi)\Big(\dot{R}H+\ddot{R}\Big)\bigg]-$$\begin{eqnarray}
m_{3}^{2}\bigg(\frac{F(R,\phi)}{F'(R,\phi)}
-R\bigg)-2V(\phi)<0.\end{eqnarray}

This is a general condition for acceleration of the universe in our
$F(R,\phi)$ DGP inspired model. It is a complicated relation and one
cannot achieve an explicit intuition of acceleration in this setup.
For simple forms of $F(R,\phi)$ we can find relatively simple
conditions that can be explained more explicitly. As a simple
example and following Faraoni [16], if we set
~$F(R,\phi)=\frac{1}{2}(1-\xi\phi^2)R$ which gives conformal
coupling of scalar field and Einstein gravity, equation (25) under
assumption of weak energy condition $\rho_{\phi}\geq0$, gives
$V-\frac{3\xi}{2}\phi\frac{dV}{d\phi}>0$, where we have assumed
$\xi\leq\frac{1}{6}$ [16]. For instance, if we set
$V(\phi)=\lambda\phi^{\eta}$, we find $ \eta<\frac{2}{3\xi}$. Using
the ansatz $a(t)\approx At^{\nu}$ and $\phi(t)\approx Bt^{-\mu}$,
equations (22) and (23) with ~$ F(R,\phi)=\frac{1}{2}(1-\xi\phi^2)R$
and $\hat{\rho}=0$, give $\xi\geq\frac{1}{12}$ with positive and
real $\nu$ and considering terms of order ~${\it O}(t^{-\mu-2})$. In
this case ~$\xi$~is restricted to the interval
~$\frac{1}{12}\leq\xi\leq\frac{1}{6}$ and since equation (23) with
these ansatz gives $\mu^2+(1-3\nu)\mu+12\xi\nu^2+6\xi\nu=0$,
positivity and reality of solutions for $\mu$ gives
$(9-48\xi)\nu^2-(6-24\xi)\nu+1\geq0$. For $\xi=\frac{1}{12}$, we
find $\nu=(4\pm\sqrt{11})/5$ which gives a power-law accelerated
expansion for positive sign. Therefore, a suitable fine-tuning of
non-minimal coupling provides the possibility of late-time
accelerated expansion. Based on a dark energy model, to have an
accelerated universe, the value of conformal NMC should be
restricted to the interval ~$0.146\leq\xi\leq 0.167$ [17]. On the
other hand, current experimental limits on the time variation of $G$
constraint the nonminimal coupling as $ - 10^{-2}\preceq \xi \preceq
10^{-2}$\,[18]. Solar system experiments such as Shapiro time delay
and deflection of light have constraint Brans-Dike parameter to be
$\omega_{BD}>500$ which leads to the result of $|\xi|\preceq
2.2\times10^{-2}$ for non-minimal coupling [18]. However, other
approaches lead to different constraints on the value of conformal
coupling [19]. As a second and more general example with $R^{n}$
correction, we set ~$
F(R,\phi)=\frac{1}{2}(1-\xi\phi^2)\ell_{0}R^n$. For spatially flat
FRW geometry the Ricci scalar is given by
\begin{equation}
R=6\frac{\ddot{a}}{a}+ 6\Big(\frac{\dot{a}}{a}\Big)^{2}.
\end{equation}
Using the above proposed ansatz for $a(t)$ and $\phi(t)$, equations
(22) and (24) can be rewritten as the following explicit
time-dependent form
$$3m_{3}^{2}\nu^2\ell_{0}n(12\nu^2-6\nu)^{n-1}\Big(1-\xi
B^2t^{-2\mu}\Big)=$$
\begin{equation}
B^2\mu^2t^{-2\mu+2n-2}+2\lambda B^\eta t^{-\mu\eta+2n}-12\xi\nu\mu
B^2\ell_{0}(12\nu^2-6\nu)^{n}t^{-2\mu-2}+m_{3}^{2}t^{2n-2}\Big((12\nu^2-6\nu)(\frac{1}{n}-1)+6\nu(n-1)\Big)
\end{equation}

$$-3m_{3}^{2}\nu(\nu-1)\ell_{0}n(12\nu^2-6\nu)^{n-1}t^{-2n+2}\Big(1-\xi
B^2t^{-2\mu}\Big)=$$ $$2B^2\mu^2t^{-2\mu}-6\xi
B^2\mu\ell_{0}(12\nu^2-6\nu)^{n}t^{-2n-2\mu}\Big(\mu+2\nu\Big)+
6\xi\ell _{0}(12\nu^2-6\nu)^{n}t^{-2n+2}\Big(\xi
B^2\ell_{0}(12\nu^2-6\nu)^{n}t^{-2n-2\mu}$$
\begin{equation}
+\lambda\eta B^{\eta}t^{-\mu\eta}\Big)+2\lambda
B^{\eta}t^{-\mu\eta+2}+6m_{3}^{2}(n-1)(2n-\nu-1)-m_{3}^{2}(\frac{1}{n}-1)(12\nu^{2}-6\nu).
\end{equation}
Our aim is to see whether these equations account for positively
accelerated expansion with suitable values of parameters. Since the
parameter space of the problem is complicated, analytical solutions
of these equations have no obvious interpretation. Therefore, we try
to see how a suitable choice of parameters leads to viability of
accelerated expansion in this setup. Our strategy is to see how with
different choices of parameters in equations (27) and (28), equality
in this equations is preserved. Firstly, we choose $\nu>1$ in the
favor of late-time positively accelerated expansion. We find
appropriate values of other parameters in this parameters space as
shown in table $1$ and $2$. We plot each side of equation (27) as a
separate function of time for some values of parameters. Existence
of intersection point for these functions shows the viability of
values attributed to parameters to preserve equality. The same
procedure for both sides of equation (28) is done too. In summary,
to have accelerated expansion, we need $\nu>1$, so choosing $\nu>1$
we find some possible values of the other parameters to have at
least one intersection point in the graph and therefore to be
consistent with accelerated expansion. Secondly, choosing some
specific values for other parameters, we obtain possible values of
$\nu$. If there is no intersection between two graphs, it means that
with that choice of parameters we will not arrive at $\nu>1$
corresponding to accelerated expansion. If there is intersection but
with $\nu<1$, the solution will be decelerating ruled out by
observational data. The results of these analysis are shown in table
$1$ and $2$ and also in figures $1$, $2$, $3$ and $4$. Note that
these are only some specific examples and the actual parameter space
is very large and complicate. However, the least advantage of this
analysis is the fact that essentially late-time acceleration can be
explained naturally within our setup. We see for instance that with
$\eta=4$, $n=1.4$, and $\xi=0.166$ there is no possibility to have
accelerated expansion with $\nu=1.2$\, since there is no
intersection point in the graph. On the other hand, accelerated
expansion with $\nu=2$ requires $\eta=2$ or $\eta=4$, $n=1.5$, and
$\xi=-0.166$ for instance. As a common role in this analysis, we
have noticed that from equation (27), only for positive values of
$n$ with $1.37\leq n<2$ and for positive or negative $\xi$ there are
intersection points and therefore accelerated expansion. These
analysis, though very simple and especial, show that $F(R,\phi)$
DGP-inspired scenarios essentially account for positively
accelerated expansion.
\begin{table}
\begin{center}
\caption{Some choices of parameters for equation (27) to be
preserved. The last column shows the coordinate of intersection
points and equality of two sides of equation (27) with corresponding
choices of parameters.} \vspace{0.5 cm}
\begin{tabular}{|c|c|c|c|c|c|c|c|}
  \hline
  \hline $\eta$ & $n$& $\nu$ & $\xi$ & $(a,b)$  \\
  \hline $2$ &$1.4$ &$1.2$& $-0.166$ &$(1.4,16.4),(7.6,15.2)$  \\
\hline $2$ &$1.4$ &$1.2$& $0.166$ &$(7.6,15.2)$  \\
\hline $4$ &$1.4$ &$1.2$& $-0.166$ &$(1.4,16.5)$  \\
\hline $4$ &$1.4$ &$1.2$& $0.166$ &$--$  \\
\hline $2$ &$1.5$ &$2$& $-0.166$ &$(1.6,114)$  \\
 \hline $2$ &$1.5$ &$2$& $0.166$ &$--$  \\
 \hline $4$ &$1.5$ &$2$& $-0.166$ &$(1.6,114)$  \\
 \hline $4$ &$1.5$ &$2$& $0.166$ &$--$  \\

 \hline
\end{tabular}
\end{center}
\end{table}

\begin{table}
\begin{center}
\caption{Some choices of parameters for equation (28) to be
preserved. The last column shows the coordinate of intersection
points and equality of two sides of equation (28) with corresponding
choice of parameters. Note that these are only some example and the
actual parameter space is very wide.} \vspace{0.5 cm}
\begin{tabular}{|c|c|c|c|c|c|c|c|}
  \hline
  \hline $\eta$ & $n$& $\nu$ & $\xi$ & $(a,b)$  \\
  \hline $2$ &$-0.37$ &$1.005$& $-0.166$ &$(63.4,46.4)$  \\
\hline $2$ &$-0.37$ &$1.005$& $0.166$ &$(122,281)$  \\
\hline $4$ &$-0.37$ &$1.005$& $-0.166$ &$(92.4,128)$  \\
\hline $4$ &$-0.37$ &$1.005$& $0.166$ &$(92.4,128)$  \\
\hline $2$ &$-0.5$ &$1.2$& $-0.166$ &$(16.9,52)$  \\
 \hline $2$ &$-0.5$ &$1.2$& $0.166$ &$(20.6,96.3)$  \\
 \hline $4$ &$-0.5$ &$1.2$& $-0.166$ &$(19,69.4)$  \\
 \hline $4$ &$-0.5$ &$1.2$& $0.166$ &$(19,69.4)$  \\

 \hline
\end{tabular}
\end{center}
\end{table}
Note that in minimal case where $ F(R,\phi)\propto R^n$ one finds
from (22) and (24) [20]
\begin{eqnarray}
\nu[\nu(n-2)+(n-1)(2n-1)]=0
\end{eqnarray}
\begin{eqnarray}
\nu[\nu(n-2)+n^2-n+1]=n(2n-1)(n-1)
\end{eqnarray}
from which the allowed solutions are
$$
\nu=0\Rightarrow n=0 ,\,\,\,n=\frac{1}{2},\,\,\,n=1
$$\begin{eqnarray} \nu=\frac{(n-1)(2n-1)}{(2-n)},\,for\,all ~n
~except ~n=2
\end{eqnarray}
Our choices of parameters in preceding discussions are based on this
argument and also constraints on non-minimal coupling from reference
[19]. Especially, we have taken into account the following
considerations: the solutions with $\nu=0$ are not interesting at
all since they provide static cosmologies with non-evolving scale
factor on the brane. In ordinary 4D gravity with the action of the
form $F(R)=\ell_{0}R^n$ and without any matter, a constant scale
factor with a singular equation of state is achieved for $n=1$.
Similarly, in the normal DGP model, that is, when $n=1$ and with no
ordinary matter present, we cannot define the equation of state
because the scale factor on the brane is constant and equation of
state parameter will be singular [10,20]. In our $F(R,\phi)$ setup
with ~$ F(R,\phi)=\frac{1}{2}(1-\xi\phi^2)\ell_{0}R^n$, as we have
discussed previously, to have accelerated expansion we need to
rstrict $n$ to the interval $1.37\leq n<2$. In this situation, there
is no term with the first power of the Ricci scalar to obtain
ordinary Einstein-like interactions between two astrophysical
objects, such as the Sun and Mercury. To overcome this problem in
our setup, we have treated the case with $n=1$ separately after
equation (25) above. On the other hand, if we choose theories such
as $F(R,\phi) = \frac{1}{2}(1-\xi\phi^{2})
\Big[R-(1-n)\zeta^{2}\Big(\frac{R}{\zeta^{2}}\Big)^{n}\Big]$ where
$\zeta$ is a suitably chosen parameter, our model will contain the
case with $n=1$ and therefore general relativity. We will come back
to this issue later with more details.
\begin{figure}[htp]
\begin{center}\includegraphics{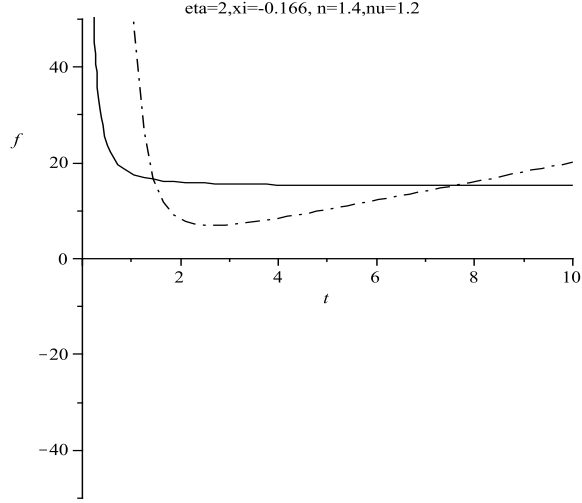} \vspace{7cm}
\end{center}
 \caption{\small {Choosing $\eta=2$, $\xi=-0.166$ and $n=1.4$, we find $\nu=1.2$ from equation (27).
 There are two intersection points accounting for positively accelerated expansion within $
F(R,\phi)=\frac{1}{2}(1-\xi\phi^2)\ell_{0}R^n$ scenario.}}
\end{figure}\\
\begin{figure}[htp]
\begin{center}\includegraphics{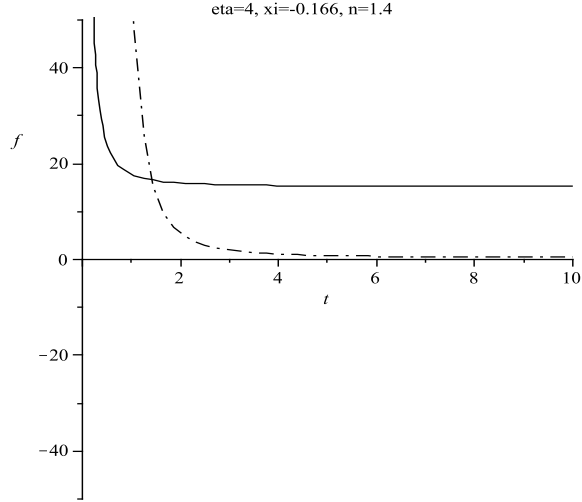} \vspace{7cm}
\end{center}
 \caption{\small {Choosing $\eta=4$, $\xi=-0.166$ and $n=1.4$, we find $\nu=1.2$ from equation (27).
 There is one intersection point accounting for equality of two sides of equation (27) and positively
 accelerated expansion with this choice of parameters.}}
\end{figure}\\
\begin{figure}[htp]
\begin{center}\includegraphics{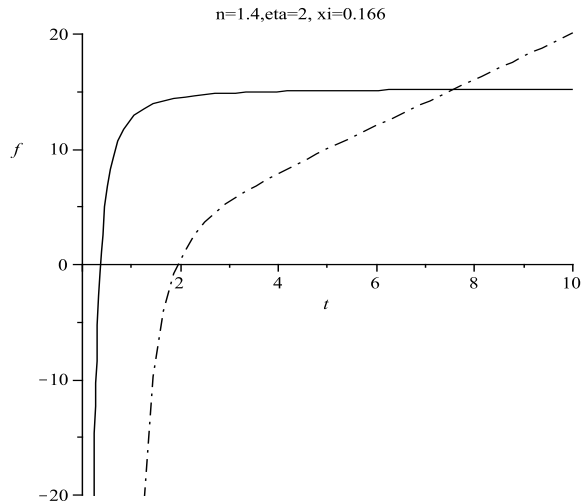} \vspace{7cm}
\end{center}
 \caption{\small { Choosing $\eta=2$, $\xi=+0.166$ ( a positive non-minimal coupling) and $n=1.4$,
 we find $\nu=1.2$ from equation (28).
 There is one intersection point accounting for equality of two sides of equation (28) and positively
 accelerated expansion with this choice of parameters.}}
\end{figure}

\begin{figure}[htp]
\begin{center}\includegraphics{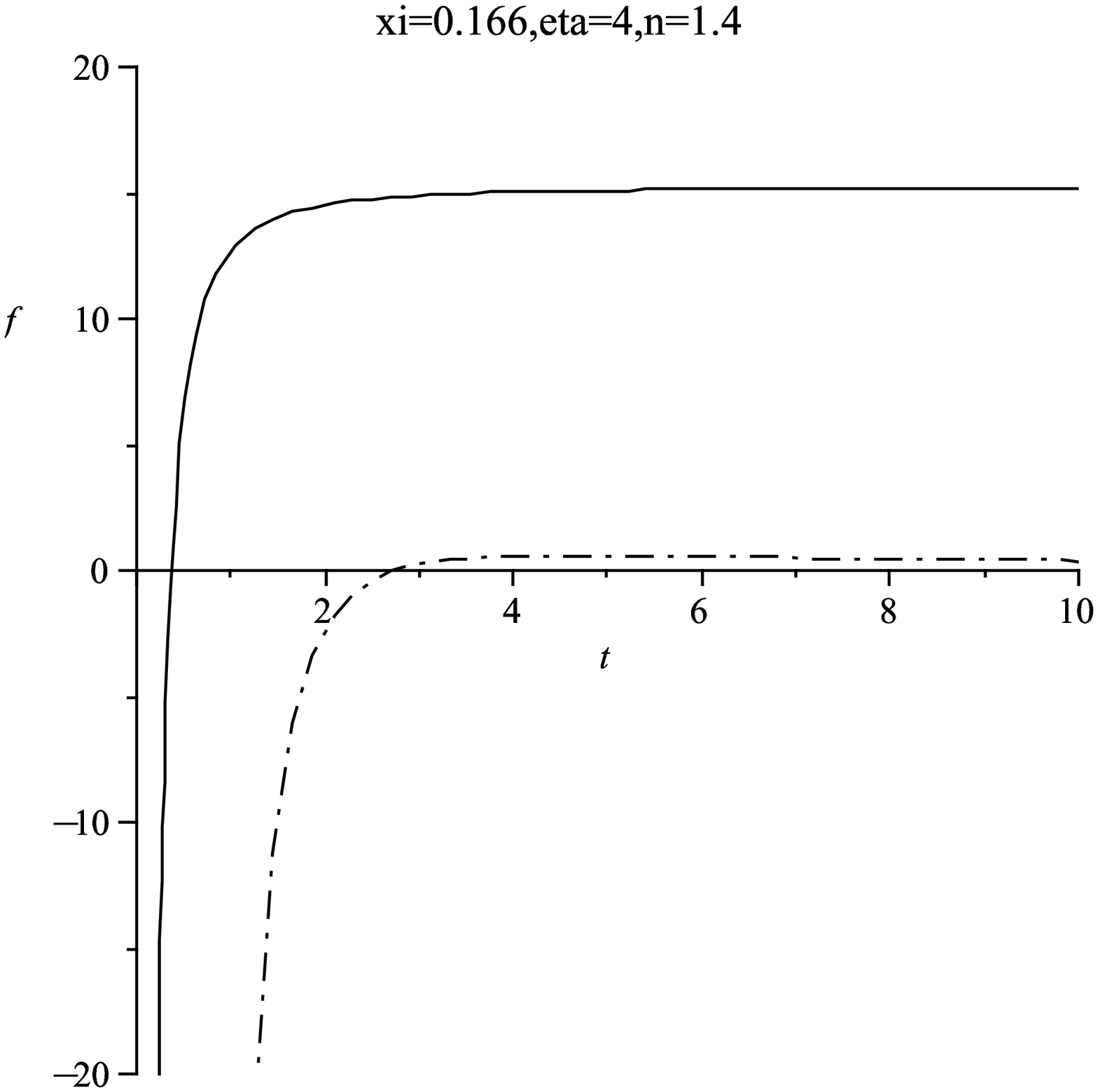} \vspace{7cm}
\end{center}
 \caption{\small { Choosing $\eta=4$, $\xi=+0.166$ and
 $n=1.4$. In this case there is no intersection point between two
 sides of equation (27) corresponding to $\nu=1.2$. It is
 interesting  to note that even with other values of $\nu>1$, there
 is no intersection point. Therefore, it is impossible to have
 accelerated expansion in this case.}}
\end{figure}

\section{Dynamics of Equation of state Parameter}
Because of explicit interaction between scalar field and curvature,
the equation of state parameter for non-minimally coupled scalar
field in this $F(R,\phi)$\, DGP-inspired setup is as follows
\begin{equation}
\omega=\frac{P^{(tot)}}{\rho^{(tot)}}=\frac{P^{(curv)}+P_{\phi}}{\rho^{(curv)}+\rho_{\phi}}.
\end{equation}
which can be written as
\begin{tiny}
\begin{eqnarray}\omega=\frac{\big(\frac{1}{2}+2\frac{d^2F}{d\phi^2}\big)\dot{\phi}^{2}-
V(\phi)+2\frac{d F}{d\phi}\big(\ddot{\phi}+2H\dot{\phi}\big)+
\frac{m_{3}^{2}}{F'(R,\phi)}\big[\big(2\dot{R}H+\ddot{R}\big)F''(R,\phi)+\dot{R}^2F'''(R,\phi)
-\frac{1}{2}F(R,\phi)+\frac{1}{2}R
F'(R,\phi)\big]}{\frac{1}{2}\dot{\phi}^{2}+V(\phi)- 6\frac{d
F}{d\phi}H\dot{\phi}+\frac{m_{3}^{2}}{F'(R,\phi)}\big(\frac{1}{2}F(R,\phi)-
\frac{1}{2}R F'(R,\phi)-3\dot{R}H\frac{d^2F}{dR^2}\big)},
\end{eqnarray}
\end{tiny}
This is a general statement for equation of state in $F(R,\phi)$
setup. To proceed further, we have to specify the functional form of
$F(R,\phi)$. Firstly we assume a conformal coupling between scalar
field and induced $R^{n}$-gravity as
$F(R,\phi)=\frac{1}{2}(1-\xi\phi^2)\ell_{0}R^n$. The scalar field
potential is chosen to be $V(\phi)=\lambda\phi^{\eta}$ and then
using natural ansatz $a(t)\approx At^{\nu}$ and $\phi(t)\approx
Bt^{-\mu}$, the following time-dependent equation of state parameter
will be derived
\begin{tiny}
$$\omega(t)=\bigg\{\mu
B^2t^{2n+\mu\eta}\Big(\frac{1}{2}t^{2n}-2\xi
\ell_{0}(12\nu^2-6\nu)^{n}(\mu+\nu)\Big)-\lambda
B^{\eta}t^{2\mu+4n+2}+2\xi \ell_{0}(12\nu^2-6\nu)^{n}t^2\Big(\xi
B^2\ell_{0}$$$$(12\nu^2-6\nu)^{n}t^{\mu\eta}+\lambda\eta
B^{\eta}t^{2\mu+2n}\Big)+m_{3}^{2}t^{2\mu+4n+\mu\eta}\Big(2(n-1)(2\nu+2n-1)
-3\nu(2\nu-1)(\frac{1}{n}-1)\Big)\bigg\}\times$$\begin{equation}\bigg\{\frac{1}{2}B^2\mu^2t^{\mu\eta+4n}+\lambda
B^{\eta}t^{2\mu+4n+2}-6\xi
B^2\mu\nu(12\nu^2-6\nu)^{n}t^{\mu\eta+2n}+3\nu(2\nu-1)(\frac{1}{n}-1)+6\nu
n(n-1)t^{2\mu+\mu\eta+4n} \bigg\}^{-1}
\end{equation}
\end{tiny}
The equation of state parameter in this scenario has a very
complicated form. To find further intuition, this equation is
plotted for several interesting cases in which follows. As these
figures show, essentially crossing of phantom divide line,
$\omega=-1$, is supported in this $F(R,\phi)$ DGP-inspired model.
However there are some points that should be stressed here. As
figure $5$ shows, the minimal case for a single scalar field has no
phantom divide line crossing for some region of parameter space. On
the other hand, this non-minimal $F(R,\phi)$ setup has no crossing
of phantom divide line with negative $n$ values. However, as table
$2$ and related arguments in preceding section show, accelerated
expansion is possible even with negative $n$ values. This is
supported by other studies [21]. For $n=-1$ we have a power-low
acceleration on the brane without having to introduce dark energy.
This result is consistent with the observational results similar to
dark energy with the equation of state parameter
$-1<\omega<-\frac{1}{3}$  which has no phantom divide line crossing.
Figures $6$ and $7$ show the time evolution of equation of state
parameter with some suitable values of model parameters.

As we have stressed previously, the case with $n=1$ needs further
discussion. We have adopted the ansatz
$F(R,\phi)=\frac{1}{2}(1-\xi\phi^2)\ell_{0}R^n$ and our analysis has
shown that $n$ should be restricted to the interval $1.37\leq n<2$
to have reliable cosmology in this setup. In this situation there is
no term with the first power of the Ricci scalar to obtain ordinary
Einstein-like interactions between two astrophysical objects, such
as the Sun and Mercury. To overcome this problem in our DGP-inspired
$F(R,\phi)$ gravity we have treated the case with $n=1$ separately
in subsection 3.1 and after equation (25). Form more general
viewpoint, we can unify these arguments by choosing the following
form of $F(R,\phi)$
\begin{equation}
F(R,\phi) = \frac{1}{2}(1-\xi\phi^{2})
\Big[R-(1-n)\zeta^{2}\Big(\frac{R}{\zeta^{2}}\Big)^{n}\Big],
\end{equation}
where $\zeta$ is a suitably chosen parameter [3,21]. The accelerated
expansion and crossing of phantom divide line in this setup can be
studied in the line of previous arguments in this paper. Fortunately
this type of theories contain $n=1$ and general relativity as a
subset. For instance, the equation of state parameter in this setup
takes the following form
\begin{tiny}
$$
\omega(t)=\bigg\{\frac{1}{2}\mu^{2}B^{2}t^{-2\mu-2}-\lambda
B^{\eta}t^{-\mu\eta}+2\xi BADt^{-\mu-2}\Big(\xi Bt^{-\mu-2}+3\mu\nu
Bt^{-\mu-2}+B^{\eta-1}\lambda\eta t^{-\mu\eta}+\mu\Big)+4\xi
B^{2}AD\mu\nu
t^{-2\mu-4}-\frac{2(1-D)}{A}\xi\mu^{2}B^{2}$$$$t^{-2\mu-2n+1}+
\frac{m_{3}^{2}A}{1+n(D-1)t^{3n-1}}\Big[
2nCA^{n-2}t^{-2n}(n-1)^{2}(-2\nu+3A^{2}+2n-4)-\frac{1}{2}Dt^{-2}\Big]\bigg\}
\times$$\begin{equation}\bigg\{\frac{1}{2}\mu^{2}B^{2}t^{-2\mu-2}+\lambda
B^{\eta}t^{-\mu\eta}-6\mu\nu\xi
B^{2}ADt^{-2\mu-4}-\frac{6m_{3}^{2}\nu
CA^{n-1}n(n-1)^{2}t^{-2n}}{1+n(D-1)t^{3n-1}}\bigg\}^{-1}
\end{equation}
\end{tiny}
where $A=12\nu^{2}-6\nu$,\, $C=\zeta^{2(1-n)}$ and
$D=1-(1-n)CA^{n-1}t^{n}$. Figure $8$ and $9$ show the result of
numerical analysis of this model. We see that these types of
theories essentially account for crossing of phantom divide line. In
this case as figure $9$ shows, we have crossing even with negative
values of $n$. Note that by choosing negative values of $n$, we can
treat theories with
$F(R,\phi)=\frac{1}{2}(1-\xi\phi^2)\Big(R-\frac{\zeta^{2(n+1)}}{R^{n}}\Big)$
in this framework. The model proposed by Carroll {\it et al} [21],
lies in this framework. For treating the issue of late time
acceleration in this case, Friedmann and acceleration equation with
this choice of $F(R,\phi)$ take the following forms
$$3\nu^{2}m_{3}^{2}D(1-\xi
B^{2}t^{-2\mu})=\mu^{2}B^{2}t^{-2\mu}+2\lambda
B^{\eta}t^{-\mu\eta+2}-12\mu\nu\xi AB^{2}t^{-2\mu-2}+12\mu\nu\xi
CA^{n}B^{2}t^{-2\mu-2n}(1-n)+$$\begin{equation}(m_{3}^{2}-1)A+\frac{12}{D}CA^{n-1}(n-1)^{2}m_{3}^{2}\nu
t^{-2n+2},
\end{equation}
and
\begin{tiny}
$$-3\nu(\nu-1)m_{3}^{2}(1-\xi
DB^{2}t^{-2\mu})=\frac{1}{2}\mu^{2}B^{2}t^{-2\mu}+\lambda
B^{\eta}t^{-\mu\eta+2}-6\mu\nu\xi
B^{2}ADt^{-2\mu-2}-6\frac{m_{3}^{2}}{D}\nu n(n-1)^{2}CA^{n-1}
t^{-2n-2}+\frac{3}{2}\mu^{2}B^{2}t^{-2\mu}-3\lambda
B^{\eta}t^{-\mu\eta+2}-$$$$6\xi B^{2}ADt^{-2\mu-2}(-\xi
DA+3\mu\nu-\lambda\eta B^{\eta-2}t^{-\mu\eta+2\mu+2})-6\mu^{2}\xi
BAt^{-2\mu-2}+6\xi\mu^{2}
CA^{n}B^2t^{-2\mu-2n}(1-n)+\frac{6}{D}\Big(-2\nu
CA^{n-1}n(n-1)^{2}t^{-2n+2}\Big)-$$\begin{equation}9A^{n-1}Cn(n-1)^{2}t^{-2n+2}-\frac{1}{4}D,
\end{equation}
\end{tiny}
where $A$, $C$ and $D$ are defined previously. Numerical analysis of
these equations shows the possibility of intersection points and
therefore accelerated expansion. For instance, figure $10$ shows the
situation for a specific choice of the parameters.

As another important point, the issue of stability of the
self-accelerated solutions should be stressed here. The
self-accelerating branch of the DGP model contains a ghost at the
linearized level [22]. Since the ghost carries negative energy
density, it leads to the instability of the spacetime. The presence
of the ghost can be attributed to the infinite volume of the
extra-dimension in DGP setup. When there are ghosts instabilities in
self-accelerating branch, it is natural to ask what are the results
of solutions decay. As a possible answer we can state that since the
normal branch solutions are ghost-free, one can think that the
self-accelerating solutions may decay into the normal branch
solutions. In fact for a given brane tension, the Hubble parameter
in the self-accelerating universe is larger than that of the normal
branch solutions. Then it is possible to have nucleation of bubbles
of the normal branch in the environment of the self-accelerating
branch solution. This is similar to the false vacuum decay in de
Sitter space. However, there are arguments against this kind of
reasoning which suggest that the self-accelerating branch does not
decay into the normal branch by forming normal branch bubbles ( see
[22] for more details). It was also shown that the introduction of
Gauss-Bonnet term in the bulk does not help to overcome this problem
[23]. In fact, it is still unclear what is the end state of the
ghost instability in self-accelerated branch of DGP inspired setups
(for more details see [22]). On the other hand, non-minimal coupling
of scalar field and induced gravity in our $F(R,\phi)$ setup
provides a new degree of freedom which requires special fine tuning
and this my provide a suitable basis to treat ghost instability. It
seems that in our model this additional degree of freedom has the
capability to provide the background for a more reliable solution to
ghost instability due to wider parameter space.

Finally, the phantom divide line crossing in conventional scalar
field theory violates positive energy theorems. In our setup,
non-minimal coupling of scalar field and induced gravity has the
capability to evade this problem. In fact, following Lue and
Starkman [24], if we consider the FRW phase instead of the
self-accelerating phase, and relax the presumption that the
cosmological constant be zero (i.e., abandon the notion of
completely replacing dark energy), then we can achieve $\omega < -1$
without violating the null-energy condition, without ghost
instabilities and without a big rip ( see [24-27] for more details).

\begin{figure}[htp]
\begin{center}\includegraphics{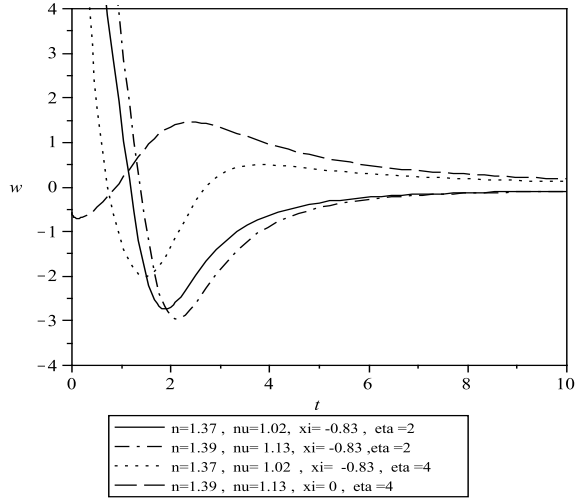} \vspace{7cm}
\end{center}
 \caption{\small {Crossing of $\omega=-1$ line in
 $F(R,\phi)=\frac{1}{2}(1-\xi\phi^2)\ell_{0}R^n$ \,\,DGP-inspired model.
 As the figure shows, with a single minimally coupled scalar field
 there are regions of parameter space with no crossing of phantom
 divide line.}}
\end{figure}

\begin{figure}[htp]
\begin{center}\includegraphics{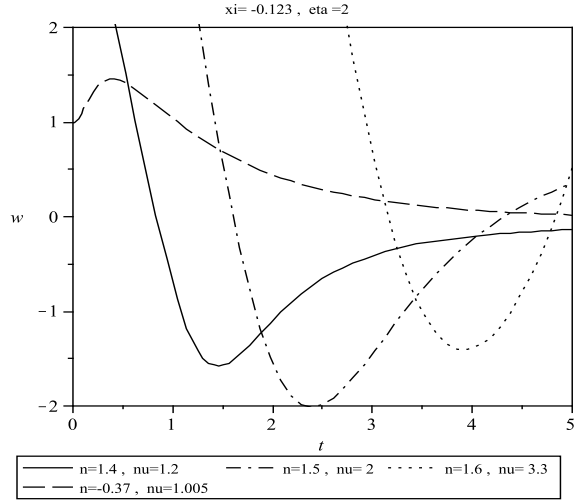} \vspace{7cm}
\end{center}
 \caption{\small {Crossing of $\omega=-1$ line. This figure shows that negative
 values of $n$ cannot account for
 phantom divide line crossing in this non-minimal setup.}}
\end{figure}

\begin{figure}[htp]
\begin{center}\includegraphics{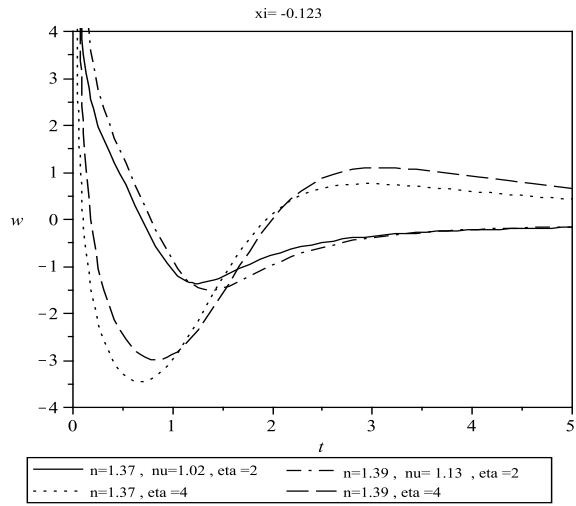} \vspace{7cm}
\end{center}
 \caption{\small {Crossing of $\omega=-1$ line in
 $F(R,\phi)=\frac{1}{2}(1-\xi\phi^2)\ell_{0}R^n$ DGP-inspired model
 for different scalar field potentials and negative non-minimal coupling.}}
\end{figure}

\begin{figure}[htp]
\begin{center}\includegraphics{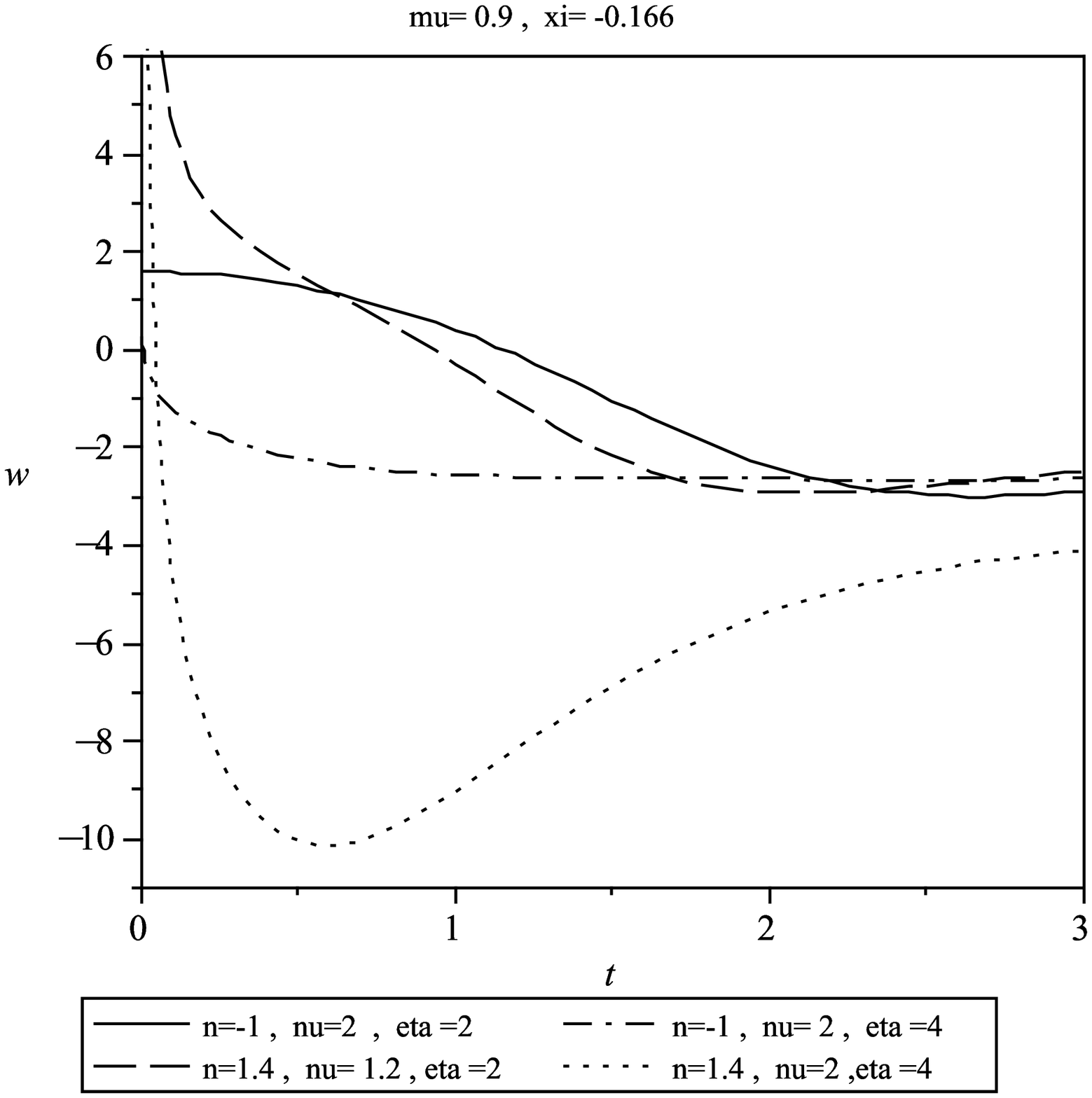} \vspace{7cm}
\end{center}
 \caption{\small {Crossing of $\omega=-1$ line in
 $F(R,\phi) = \frac{1}{2}(1-\xi\phi^{2})
\Big[R-(1-n)\zeta^{2}\Big(\frac{R}{\zeta^{2}}\Big)^{n}\Big]$
DGP-inspired model
 for different scalar field potentials and negative non-minimal coupling.}}
\end{figure}

\begin{figure}[htp]
\begin{center}\includegraphics{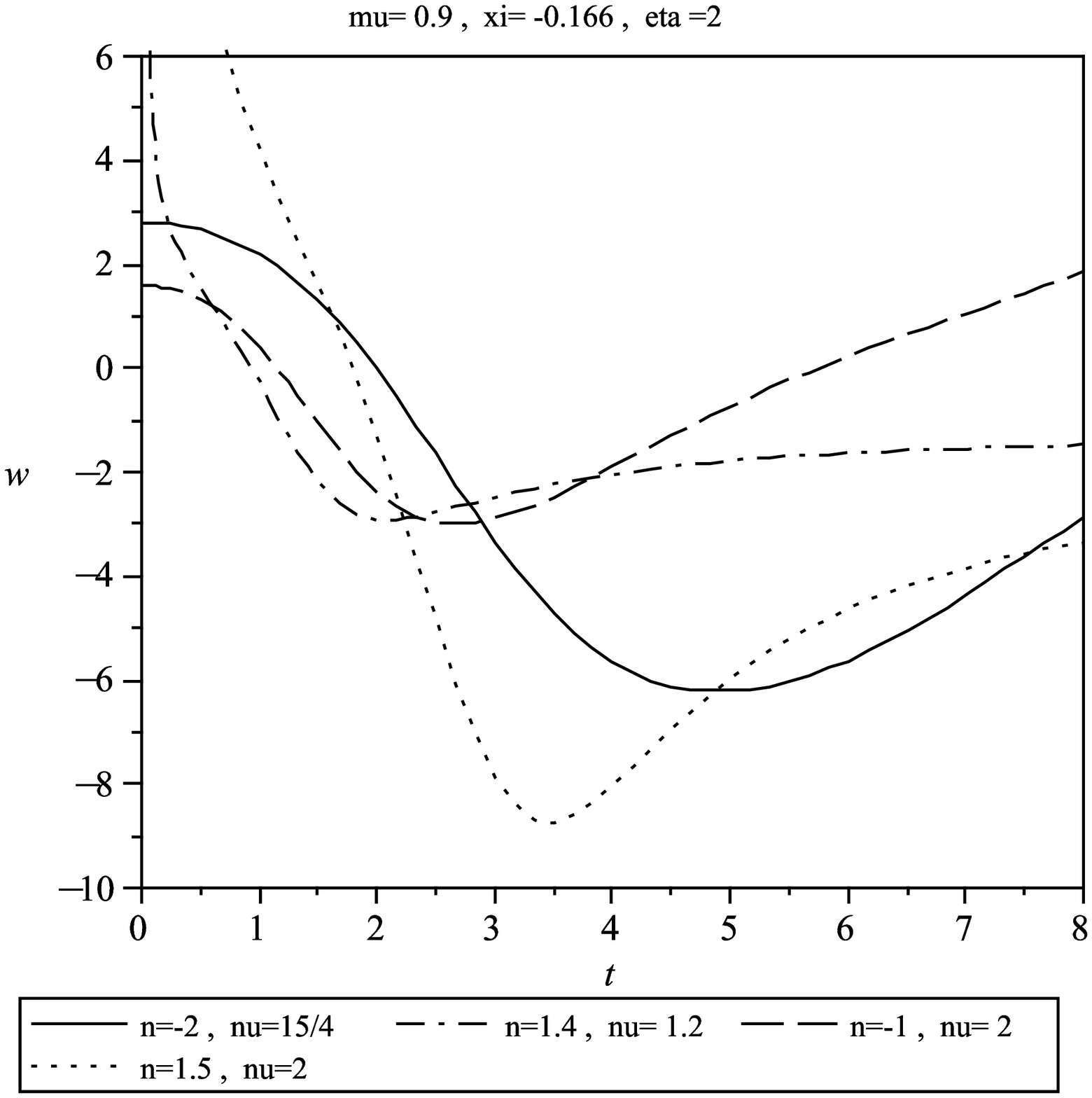} \vspace{7cm}
\end{center}
 \caption{\small {Crossing of $\omega=-1$ line in
 $F(R,\phi) = \frac{1}{2}(1-\xi\phi^{2})
\Big[R-(1-n)\zeta^{2}\Big(\frac{R}{\zeta^{2}}\Big)^{n}\Big]$
DGP-inspired model
 for different $n$ and negative non-minimal coupling. }}
\end{figure}

\begin{figure}[htp]
\begin{center}\includegraphics{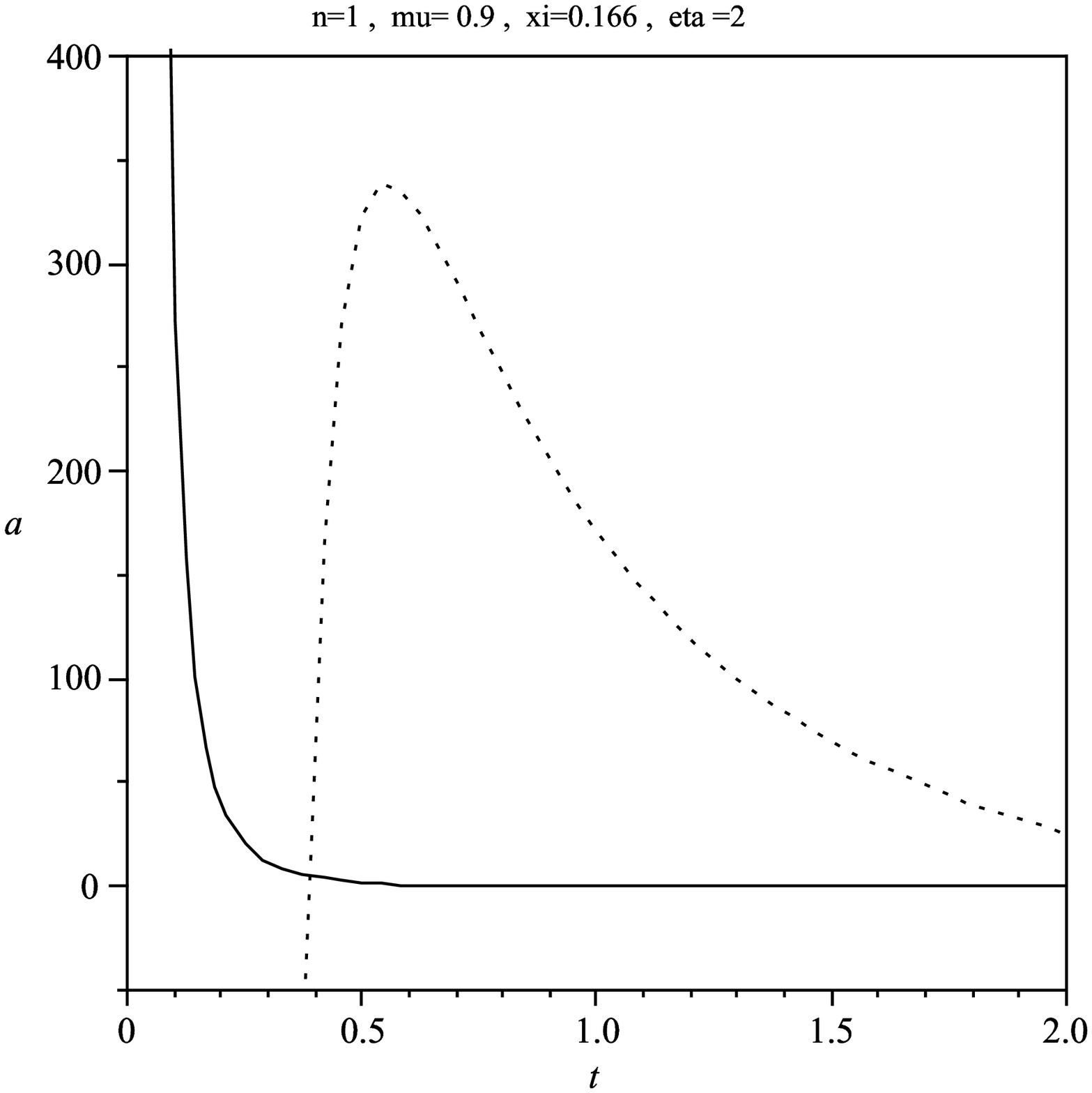} \vspace{7cm}
\end{center}
 \caption{\small {With $\eta=2$, $\xi=0.166$ and $n=1$, we find $\nu=1.2$ from equation (38).
 There is an intersection point between curves corresponding to two sides of this equation
 with $\nu=1.2$ accounting for positively accelerated expansion within
 $F(R,\phi) = \frac{1}{2}(1-\xi\phi^{2})
\Big[R-(1-n)\zeta^{2}\Big(\frac{R}{\zeta^{2}}\Big)^{n}\Big]$
scenario. }}
\end{figure}
\section{Summary and Conclusions}
Current positively accelerated expansion of the universe could be
the result of a modification to the Einstein-Hilbert action. Also
DGP braneworld scenario has the capability to interpret this
late-time acceleration via leakage of gravity to extra dimension in
its self-accelerating branch. On the other hand, there are several
compelling reasons for inclusion of an explicit non-minimal coupling
of scalar field and induced gravity on the brane sector of DGP
action [14,16]. These arguments have led us to study cosmological
implications of a DGP-inspired $F(R,\phi)$ gravity. Although this
issue has been studied extensively for the minimal case, our setup
provides a new approach to treat accelerated expansion within a
general framework of {\it modified} scalar-tensor theories taking
into account the role played by curvature correction and non-minimal
coupling of scalar field and modified curvature simultaneously. We
have studied late-time acceleration and possible crossing of phantom
divide line in this setup. The condition for accelerated expansion
and also equation governing on the dynamics of equation of state
parameter are very complicated, so that we were forced to try a
reliable and natural ansatz to find some intuition. We have shown
that this setup accounts for accelerated expansion with a suitable
choice of parameters space or fine-tuning. Also, this setup accounts
for crossing of phantom divide line in some ranges of parameters
appeared in the model. In the minimal case, our model coincides with
existing models of $f(R)$ gravity. Especially the case with $n=-1$
in the absence of non-minimal coupling exactly coincides with
well-known result [21]. On the other hand, for a single scalar field
with minimal coupling to gravity there are region of parameter space
with no phantom divide line crossing. We have seen in the numerical
calculations that with $V(\phi)=\lambda \phi^{\eta}$, for both
$\eta=2$ and $\eta=4$ and with arbitrary $\xi$, there is no crossing
of phantom divide line for negative values of $n$ ( Figure $6$ ).
However positively accelerated expansion of the universe can be
explained in this situation with negative $n$. On the other hand, if
$1.37\leq n<2$, for negative $\xi$ we have crossing of the phantom
divide line ( see figure $7$ ). As an important result in the
analysis of late-time behavior, from equation (27) we find that only
for positive values of $n$ with $1.37\leq n<2$ and for positive or
negative values of $\xi$, there are intersection points and
therefore accelerated expansion for corresponding parameters choice.
The case with $n=1$ which describes ordinary general relativity has
been treated separately. By adopting more general theories such as
$F(R,\phi) = \frac{1}{2}(1-\xi\phi^{2})
\Big[R-(1-n)\zeta^{2}\Big(\frac{R}{\zeta^{2}}\Big)^{n}\Big]$, the
previous restriction on $n$ is relaxed so that $n=1$ is a natural
subset of the model. In summary, accelerated expansion can be
explained by both positive and negative values of $n$ with suitable
fine-tuning of other parameters. On the other hand crossing of
phantom divide line in ~$
F(R,\phi)=\frac{1}{2}(1-\xi\phi^2)\ell_{0}R^n$ model can be
explained only with positive values of $n$ and fine-tuning of other
parameters. The range of variation of $n$ to have crossing of
phantom divide line in this setup is $1.37\leq n<2$. However in
theories with $F(R,\phi) = \frac{1}{2}(1-\xi\phi^{2})
\Big[R-(1-n)\zeta^{2}\Big(\frac{R}{\zeta^{2}}\Big)^{n}\Big]$ this
crossing occurs for both negative and positive values of $n$.  For
theories of the type $F(R,\phi) = \frac{1}{2}(1-\xi\phi^{2})
\Big[R-(1-n)\zeta^{2}\Big(\frac{R}{\zeta^{2}}\Big)^{n}\Big]$, $n$ is
not restricted to this interval and these theories account for
phantom divide line crossing even for negative values of $n$ as
figure $9$ shows. The issues of ghost instabilities and violation of
positive energy theorems have been discussed also. Due to wider
parameter space as a result of non-minimal coupling of scalar field
and modified gravity, we hope this model can evade these problems.\\

{\bf Acknowledgment} We are indebted to an anonymous referee for
his/her important contribution in this work.

\end{document}